\documentclass[RNAAS]{aastex62}

\usepackage{amsmath, amssymb, bm}

\begin{document}

\title{Retrieving Internal Kinematics of Galaxies with Deep Learning using Single-Band Optical Images}

\correspondingauthor{Christopher J. Conselice}
\email{conselice@manchester.ac.uk}

\author{Sakina Hansen}
\affiliation{Centre for Astronomy and Particle Theory, University of Nottingham, Nottingham NG7 2RD, United Kingdom}
\author{Christopher J. Conselice}
\affiliation{Centre for Astronomy and Particle Theory, University of Nottingham, Nottingham NG7 2RD, United Kingdom}
\affiliation{Jodrell Bank Centre for Astrophysics, University of Manchester, Oxford Road, Manchester UK}
\author{Amelia Fraser-McKelvie}
\affiliation{Centre for Astronomy and Particle Theory, University of Nottingham, Nottingham NG7 2RD, United Kingdom}
\affiliation{International Centre for Radio Astronomy Research, The University of Western Australia, 35 Stirling Hw, 6009 Crawley, WA, Australia}
\author{Leonardo Ferreira}
\affiliation{Centre for Astronomy and Particle Theory, University of Nottingham, Nottingham NG7 2RD, United Kingdom}


\begin{abstract}

Using deep machine learning we show that the internal velocities of galaxies can be retrieved from optical images trained using 4596 systems observed with the SDSS-MaNGA survey.  Using only $i$-band images we show that the velocity dispersions and the rotational velocities of galaxies can be measured to an accuracy of 29 km~$\rm{s}^{-1}$ and 69 km~$\rm{s}^{-1}$ respectively, close to the resolution limit of the spectroscopic data.   This shows that galaxy structures in the optical holds important information concerning the internal properties of galaxies, and that the internal kinematics of galaxies are quantitatively reflected in their stellar light distributions beyond a simple rotational vs. dispersion distinction.

\end{abstract}

\vspace{-0.1cm}
\section*{Introduction} 

Galaxy structure has remained one of the major features of galaxies which contains  important physical information we are still uncovering.  For example, soon after the galaxy classification of Hubble, it was shown by e.g., Roberts (1969) that different Hubble types have different properties - namely that ellipticals are red and low star-forming, and that spirals are blue and show evidence for ongoing formation.
Since that time there has remained an outstanding question about whether or not the structure of a galaxy gives clues to its internal properties and formation history.  We know that star-formation is proportional to the clumpiness of light in a galaxy, mergers correlate with asymmetry, and the stellar mass to the concentration of the light (e.g., Conselice 2003; Graham et al. 2001). In general, more concentrated galaxies have a higher mass and larger internal velocities.

One of the issues with this, however, is that the correlation with internal properties is based on measuring a structural parameter derived from the structure itself.  What we would also like to know is if galaxy structure reveals internal kinematics. That is, is the distribution of light affected by the velocity dispersion or rotational velocity, and vice-versa?

Presented in this work is a deep machine learning regression model which aims to answer this question and automates the fast determination of velocity dispersions and rotational velocities of galaxies from imaging. A convolutional neural network is used to extract features from images from the Sloan Digital Sky Survey (SDSS) database, alongside spectroscopic data of these galaxies obtained from the Mapping Nearby Galaxies at Apache Point Observatory (MaNGA) survey.

\vspace{0.1cm}
\section*{Data Set and Method}

The spectroscopic data we use were observed by the MaNGA galaxy survey (Bundy et al. 2015), which is an SDSS-IV project. SDSS data release 15 (Aguado et al. 2019) contains MaNGA data with integrated field spectroscopic (IFS) observations of 4596 unique galaxies with continuous wavelength coverage between 3600--10300 \AA~ at a resolution of R$\sim2700$ ($\sigma\sim70~\rm{km}~\rm{s}^{-1}$). MaNGA's Primary+ sample contains observations out to $\sim1.5~R_{e}$ for $\sim$66\% of the sample, and the Secondary sample has observations out to $2.5~R_{e}$ for the remainder of the targets.   The MaNGA galaxies cover a wide range of stellar masses $\sim10^{7}~\rm{M}_{\odot}<M_{\star}< \ \sim10^{12}~\rm{M}_{\odot}$.

Morphologies are provided for every galaxy in the MaNGA sample from Galaxy Zoo 2 (GZ2) classifications (Willett et al. 2013). We separate early- and late-type galaxies using the weighted GZ2 vote fractions and classify late-type galaxies as those with \texttt{p$\_$spiral$\_$weighted $>$ 0.6} and early-types as those with \texttt{p$\_$smooth$\_$or$\_$features $>$ 0.6}.

Stellar rotational velocity and velocity dispersion ($\sigma$) maps are obtained via the \textsc{Marvin} \textsc{Python} tool (Cherinka et al. 2019), while correcting the $\sigma$ maps to the astrophysical dispersion (Westfall et al. 2019). These maps were produced by the MaNGA data analysis pipeline (Westfall et al. 2019) and we extract the maximum rotational velocity within the aperture, and the maximum $\sigma$.   These maps were used in conjunction with SDSS $i$-band imaging to determine the link between galaxy morphology and internal kinematics.  These are then analyzed with a deep learning methodology to determine through using a Convolutional Neural Network (CNN) (e.g., Cheng et al. 2020) the best regression possible.

For our CNN, the input data we use are the 4596 SDSS images and their matching kinematic data -- the MaNGA velocities. We split our sample into a `training' set and a `test' set in an 85\% / 15\% split respectively, i.e. 3,906 galaxies for training and 690 galaxies to test the network. The sample images have their flux normalised and are re-scaled to an input pixel resolution of 64 by 64, with a single channel for gray-scale images.  Normalisation is important so that the values in the input images are in the appropriate range of the network activation functions.
 
Our model works by pattern recognition, i.e. discovering meaningful correlations between the labels and the images in the training set. The model then tests what it has `learnt' from the training set on the test set.  Overall the model consists of two parts. The first part consists of convolutional layers that extract features from the input images using convolution operations, while the second part of the model combines these features together in fully connected layers of neurons to produce the final prediction.  Max Pooling layers reduce the input data scale so the following convolutional layers extract features in different scales representation by discarding irrelevant information. 
 Data augmentation is used to increase the amount of training data for the purpose of increasing generalizations.
Examples of data augmentation we use include rotating, zooming, shifting and flipping images.  The loss function used is the mean absolute error.

 The success of the model is measured by how accurately it predicts the true values. As a regression predictive model is used, the training step goal is to minimize a loss function that estimate the error in those predictions.  The network is trained until convergence, which is through 100 `epochs'. During each epoch, the weights and convolutional kernels are updated and the model is re-evaluated. 
 

\vspace{0.1cm}
\section*{Results}

Our regression model finds a correlation between imaging and velocity dispersion which reaches a mean absolute error of 29 km $\rm{s}^{-1}$ (20\% error). For rotational velocity, the results were not as conclusive, with the regression model reaching a mean absolute error of 69 km $\rm{s}^{-1}$ (43\% error). The results of our CNN regression showing the model predictions on the validation data are shown in Figure~1.  The best fit for dispersion is given by:


\begin{figure}[h]
\begin{center}
\includegraphics[scale=0.4,angle=0]{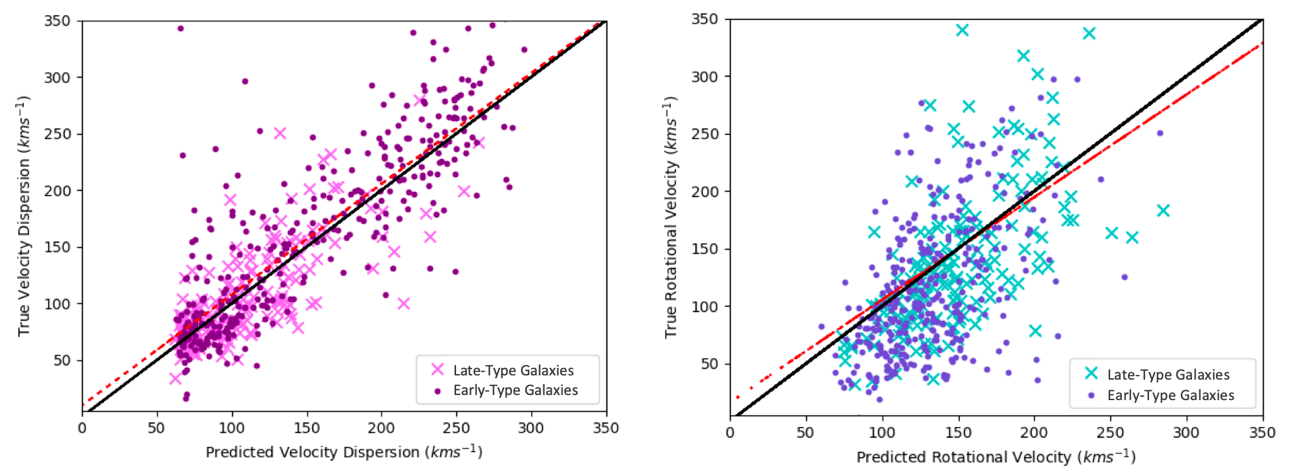}
\caption{The predicted vs. measured internal velocities for galaxies in test our sample. The x-axis gives the predicted velocity dispersion and rotational velocity, and the y-axis are the actual values. The red dashed line is the fit, and the black line indicates the 1:1 relation.}
\end{center}
\end{figure}

\begin{equation}
    \sigma_{actual} = 0.98 \times \sigma_{predict} + 9.55
\end{equation}

\noindent where the scatter amongst this fit is $\sim 29$ km s$^{-1}$.  For the rotational velocity we find that the best fit is given by:

\begin{equation}
    V_{actual} = 0.94 \times V_{predict} + 16.55
\end{equation}

\noindent where the scatter on this value is $\sim 64$ km s$^{-1}$.   Both of these values are similar to the resolution limit of the MaNGA data (e.g., Bundy et al. 2015).  This means that with just a single imaging band we are now able to retrieve kinematic information for galaxies.  This also shows that kinematic information, such as the level of rotation and internal velocities, are reflected in the structure of a galaxy and the distribution of its stars.  It remains to be seen if the structure correlates furthermore with the total mass which would involve investigating the combination of size with these internal velocities.  


Overall, we can conclusively say that the optical image of a galaxy  contains information concerning the kinematics of the system, in addition to the merger state already published (Ferreira et al. 2020).


\vspace{0.1cm}
\section*{Implications}

We find a stronger correlation exists between morphology and velocity dispersion than between the rotational velocity.  It is possible that we could improve the fit for the rotational velocity with other information, such as images in other wavelengths.   The increased scatter in the morphology-rotational velocity relation is likely due to the difficultly in separating the rotation-dominated lenticular galaxies from dispersion-dominated ellipticals using the crude GZ2 cut we employ.   As these types cannot be separated without the kinematics of the galaxy, this is likely partially the cause of the scatter in the correlations.

Another explanation of the difference in correlation for the velocity dispersion and the maximum rotational velocity could be due to the aperture size of the MaNGA observations. 66\% of our galaxies have a coverage out to 1.5$R_{e}$, and the remainder out to 2.5$R_{e}$. We expect that not all galaxies will reach the maximum value of their velocity curves by 1.5$R_{e}$, resulting in measured rotational velocities that are lower than their true maximum. As the maximum velocity dispersion occurs in the centre of the galaxy, aperture does not cause a problem for $\sigma$ measurements. 

There are two main implications for these results, one which is scientific and one which is practical. First, it is still unknown what type of information a galaxy's structure contains.  Does galaxy structure somehow reflect the internal properties of galaxies, such as its physical state or formation history (e.g., Conselice 2006), or is it a semi-random and transient feature analogous to weather? Our results strongly suggest the former;  galaxy structure is a reflection of the kinematics and dynamics of galaxies.  Second,  we can now measure basic kinematic properties of galaxies using just images, which can  be used to create large catalogs of galaxy kinematics, which otherwise are difficult and expensive to obtain.




\end{document}